\documentclass[10pt, journal]{IEEEtran}
\usepackage{graphicx}
\usepackage{subfigure}
\usepackage{epstopdf}
\usepackage[compress]{cite}
\usepackage{amsfonts,amssymb}
\usepackage{amsmath}
\usepackage{changepage}
\usepackage{algorithm}
\usepackage{algpseudocode}
\usepackage{array}
\usepackage{multirow}

\usepackage{threeparttable}

\makeatletter
\def\ps@headings{%
\def\@oddhead{\mbox{}\scriptsize\rightmark \hfil \thepage}%
\def\@evenhead{\scriptsize\thepage \hfil \leftmark\mbox{}}%
\def\@oddfoot{}%
\def\@evenfoot{}}
\makeatother
\renewcommand{\baselinestretch}{1.0}

\begin{document}
\title{\LARGE RF Energy Harvesting Sensor Networks for Healthcare of Animals: Opportunities and Challenges}
\author{Yu Luo, Lina Pu, Yanxiao Zhao\\Electrical \& Computer Engineering,
South Dakota School of Mines and Technology, SD, USA 57701\\
Email: \{yu.luo, lina.pu, yanxiao.zhao\}@sdsmt.edu
}

\maketitle

\begin{abstract}
\label{sec:abstract}
In recent years, the radio frequency (RF) energy harvesting technique is considered as a favorable alternative to supply power for the next-generation wireless sensor networks. Due to the features of energy self-sustainability and long lifetime, the energy harvesting sensor network (RF-EHSN) becomes a promising solution to provide smart healthcare services. Nowadays, many energy harvesting based applications have been developed for monitoring the health status of human beings; how to benefit animals, however, has not yet drawn people's attention. This article explores the potential of applying RF-EHSNs to monitoring the health level of animals. The unique challenges and potential solutions at different layers of an RF-EHSN for animals' healthcare service are studied.
 \end{abstract}

\section{Introduction}
\label{sec:introduction}

Harvesting energy from radio environment has been considered as a promising technique to drive the next-generation wireless sensor networks~\cite{shaikh2016energy}. Not only is RF energy harvesting self-sustaining and pollution-free, but an added advantage is its ability to enable perpetual monitoring in a wide spectrum of applications (e.g., healthcare, emergency management, and fitness). Recently, considerable efforts have been made to provide smart health services for human beings using the radio frequency energy harvesting sensor networks (RF-EHSNs)~\cite{hu2017wireless}; the pets, livestock and wildlife, however, have received limited benefits from this favorable technique.

Studies reveal that animals usually hide their illness and pretend to be well even when they are sick. This masking comes from animals' instinct for survival. An obvious example would be the case of a predator who targets the weakest member of a group during a hunt. This masking feature makes it difficult for veterinarians to save animals' lives once clear illness symptoms appear. Therefore, it is important to frequently check animals' health status and behaviors for a timely treatment.

However, it is highly inefficient to hire manpower to examine the health of animals, especially when the number is large. For instance, the Noble Foods Ltd is the largest free-range egg supplier in the U.K. It owns around 100 farms with an average flock size of 5,000-6,000. How to monitor the health of so many hens for disease control is a critical problem. Another example is the Owens Aviary in the San Diego Zoo, which is the home of around 200 tropical birds. It is almost impossible to monitor the health status of each bird in such a large aviary.

The wearable RF-EHSN provides a cost-effective and affordable solution to monitor the health of pets and animals. Current biosensors developed for animals have been able to conduct physiological measurement (e.g., body temperature, heartbeat, and respiratory rate), behavior tracking, and pathogen detection~\cite{neethirajan2017recent}, and provide credible information about health conditions of the carrier. By ``capturing'' renewable RF energy from surrounding environment, an energy harvesting node (EHN) can perform sensing and communication for a long period of time without battery replacement; which is an attractive advantage to provide the health service for a large number of animals in a wide area.

Current research on RF-EHSN based smart healthcare, such as the body area network (BAN), wearable internet of things (IoT), and implantable biomedical devices, mainly focus on the human beings~\cite{salayma2017wireless, sun2016edgeiot}. Limited knowledge is known about the performance of applying an RF-EHSN into the health monitoring of animals and no attempt has been made in this research direction. To fill the gap, we attempt to investigate this new and exciting area, which can have a significant impact on disease prevention and behavioral surveillance for animals in farms and zoos.

This article gives a tutorial on the healthcare of animals with the RF-EHSN technology. A multi-tier architecture is proposed to provide long-range communications between biosensors and a health center. Afterward, the opportunity of applying the fifth-generation (5G) mobile technology to replenish EHNs for high-speed communications is discussed. Compared with conventional wireless sensor networks, RF-ENSN has some unique features. For instance, the data packet from an EHN not only carries information for communication but also contains energy that can charge neighboring EHNs; in addition to the data transmissions, EHNs also require to reserve the time and channel for the energy harvesting process; the relative positions of implanted biosensors change repetitively with the body movement of an animal. How those features challenge the design of an RF-ENSN at the physical layer, link layer, and network layer is analyzed carefully. Eventually, some potential solutions that utilize the unique features of RF-ENSN to improve the quality of health service for animals are introduced.

\section{Network Architecture}
\label{sec:arch}

\begin{figure}[htb]
\centerline{\includegraphics[width=8.8cm]{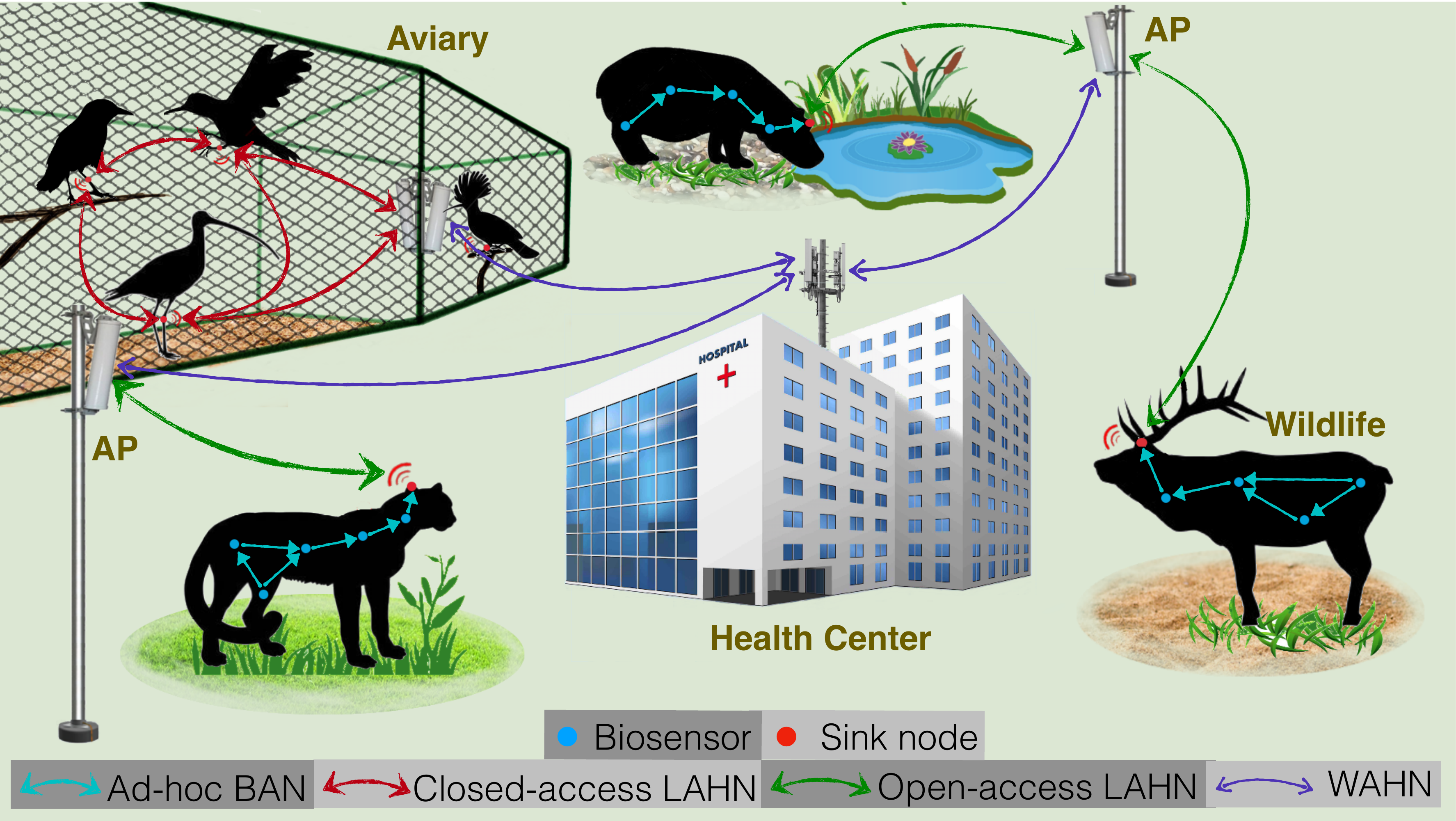}}
  \caption{The 3-tier architecture of RF-EHSNs for health monitoring of animals in a zoo.}\label{fig:scenario}
\end{figure}

Due to the size and the energy constraint of the wearable EHNs, their communication range is short (usually less than one feet)~\cite{kurup2012body}. We propose a 3-tier network architecture to deliver the sensing data collected from animals in a large area to a far located health center, as illustrated in Fig.\,\ref{fig:scenario}.

In the Tier 1 RF-EHSN, EHNs that carry biosensors implanted inside animals' body form an ad-hoc body area network (BAN). Each BAN is associated with a sink node, which is a special EHN embedded in a necklet, earring, or leg band of an animal. It is exposed in vitro and thus can have a relatively larger size and better channel quality for mid-range (hundreds of feet) data communications.

After gathering information from the associated BANs, sink nodes forwards the data to an access point (AP), forming a local area healthcare network (LAHN), which is the Tier 2 RF-EHSN. For animals like birds that live in a hutch or an aviary enclosing with an iron net, a Faraday cage is formed and blocks the electromagnetic waves. In this circumstance, an AP needs to be placed inside the cage acting as a dedicated energy source to power EHNs; the corresponding network is a closed-access LAHN. If APs are deployed in an open area, such as a wild animal zoo or a dairy farm, an open-access LAHN is established, where EHNs can harvest energy from both ambient RF environment and APs.

Eventually, a wide area healthcare network (WAHN), the Tier 3 RF-EHSN, is developed by connecting the APs in LAHNs to Ethernet. The operator in a health center can process the information sent from APs to monitor the health status of animals in real-time. This 3-tier RF-EHSN enables a timely diagnosis to prevent diseases like bird flu and mad cow disease, thereby saving the lives of animals for dairy cattle, poultry farms, and wildlife parks.

\section{RF-EHSN with 5G technology }
\label{sec:5G}
In this section, we study the potential of applying 5G technology into RF-EHSN for the animals' health monitoring. According to the measurement reported in \cite{pinuela2013ambient}, the density of ambient RF signals in urban and semi-urban environments is between $0.18$ and $84$\, nW/cm$^2$ on DTV, GSM 900/1800, 3G and WiFi frequencies on average, which is considerably thin to replenish an EHN in a short period of time.

The 5G mobile technology provides an opportunity to solve the thin energy problem. By operating in the millimeter wave (mmWave) band, the size of a transceiver can be significantly reduced, which makes a dense deployment of femtocells with massive antennas possible~\cite{buzzi2016survey}. By utilizing the femtocell as APs in an RF-EHSN, a highly directional beamforming can be steered to the direction of the desired EHN, which increases the directive gain and enables a quick energy charge.

The challenge of directional charging is the tight time synchronization across the antennas of a femtocell and accurate channel state information (CSI) between EHN and femtocell. These requirements can be achieved with the assistance of positioning technologies in 5G mobile networks, which can provide the location and energy status of active femtocells to EHNs. With such information, an EHN can receive RF signals from the most appropriate femtocell at the right time to optimize the efficiency of energy charge.

Compared with passive energy harvesting from ambient RF signals, an RF-EHSN with the 5G technology allows EHNs to request energy from femtocells on-demand. The advantage of proactive energy requests is the improved efficiency of power management. Specifically, the majority of existing transmission scheduling in RF-EHSN works in an offline scenario, assuming that the amount of energy harvested by an EHN at a given time is known in advance~\cite{ulukus2015energy}. Such assumption may not be true in the ambient RF environment, where the amount and the time of energy arrival are both random. By contrast, requesting energy from femtocell actively can guarantee that the energy received by an EHN is controllable and predictable to a certain extent, thereby making offline strategies more realistic in real applications.

When supplying energy to RF-EHSNs with 5G technology, the sensitivity of mmWave to the blockage needs to be taken into account. RF signals have weak ability to diffract around obstacles of sizes significantly larger than the wavelength. For the EHN implanted inside a moving animal, the mmWave link will become intermittent. Therefore, maintaining a reliable connection between an RF-EHSN and femtocells in a 5G network is a challenge for both energy charging and data transmission in animal healthcare applications.

\section{Challenges at physical layer}
\label{sec:PHY}
Although energy harvesting can provide perpetual energy, it is critical to intelligently utilize harvested energy to achieve optimal networking performance and boost energy utilization efficiency. In this section, we identify the challenges at the physical layer of RF-EHSN and propose a new feedback-based energy harvesting model. 

\begin{figure}[htb]
\centerline{\includegraphics[width=8.3cm]{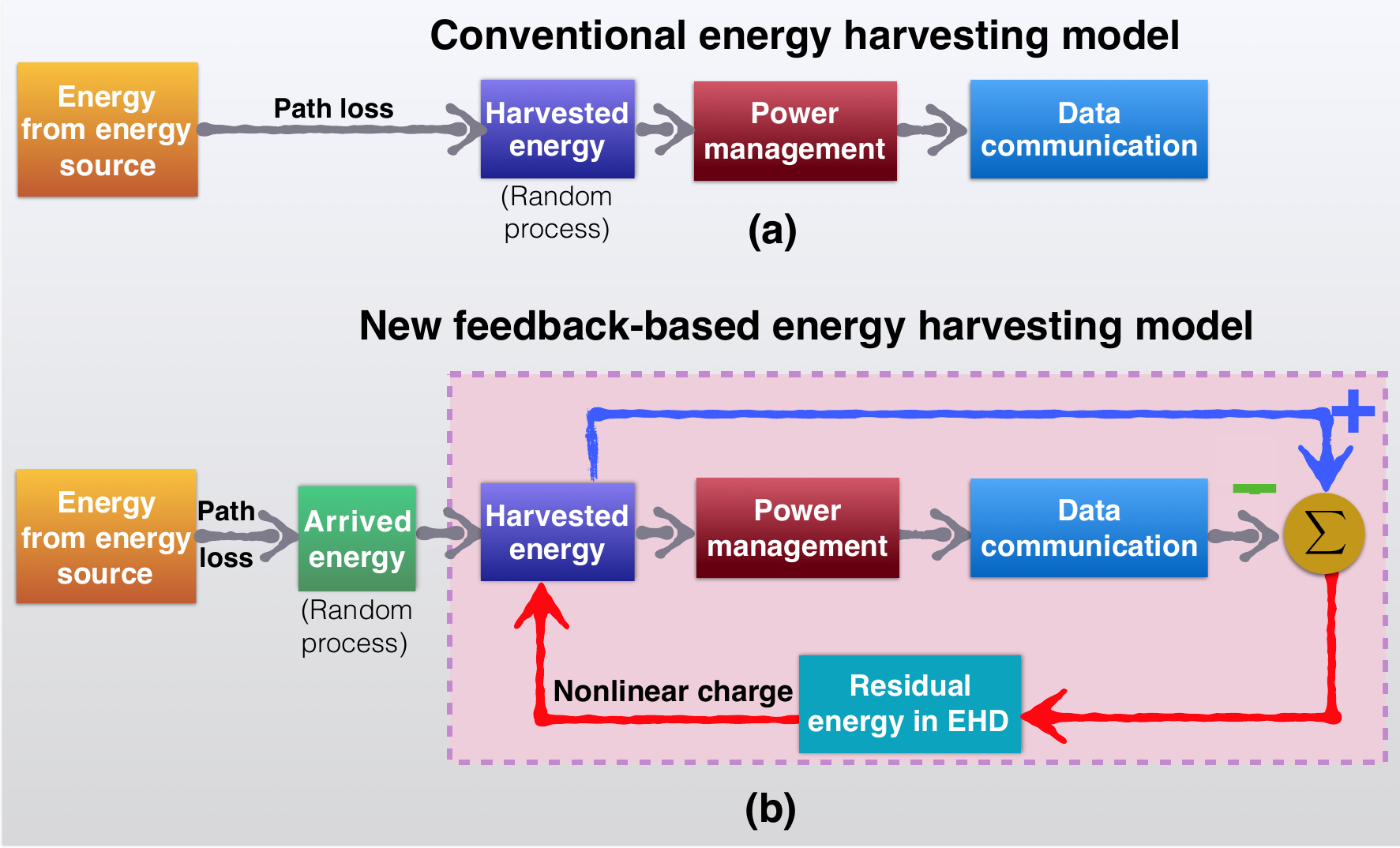}}
  \caption{Energy harvesting models before and after taking the nonlinear charge characteristic of an EHN into account: (a) The conventional one, and (b) the new feedback-based model.}\label{fig:Mod}
\end{figure}

Fig.~\!\ref{fig:Mod}\,(a) demonstrates the conventional energy harvesting model, where the amount of harvested energy is modeled as an independent random process. Although the dependency of energy harvesting efficiency on the concurrent charge state (i.e., residual energy) of batteries is a well-known fact~\cite{boshkovska2015practical, biason2016effects}, it is not taken into account in the power management on EHNs. It was suggested that EHNs would harvest an equivalent amount of energy as long as the arrived energy is the same, regardless of the residual energy on EHNs~\cite{ulukus2015energy}. 

Unfortunately, this assumption may not be true in the real world. Due to the nonlinear charge characteristic of batteries, the harvested energy is also significantly affected by the residual energy of an EHN, as illustrated in Fig.~\!\ref{fig:Mod}\,(b). To further evaluate the accuracy of the nonlinear charging feature, an indoor experiment is conducted using Powercast P2110 development kit, and Micro850 programmable logical controller (PLC). In the test, an EHN is scheduled to receive RF energy radiated from a dedicated ES, TX91501 transmitter, which was placed 6 feet away. A supercapacitor with a series resistance is selected as the battery. Since the TX91501 is non-programmable, the on/off time of the transmitter is manipulated via the Micro850 controller to direct the charging time, $T$.

\begin{figure}[htb]
\centerline{\includegraphics[width=7.7cm]{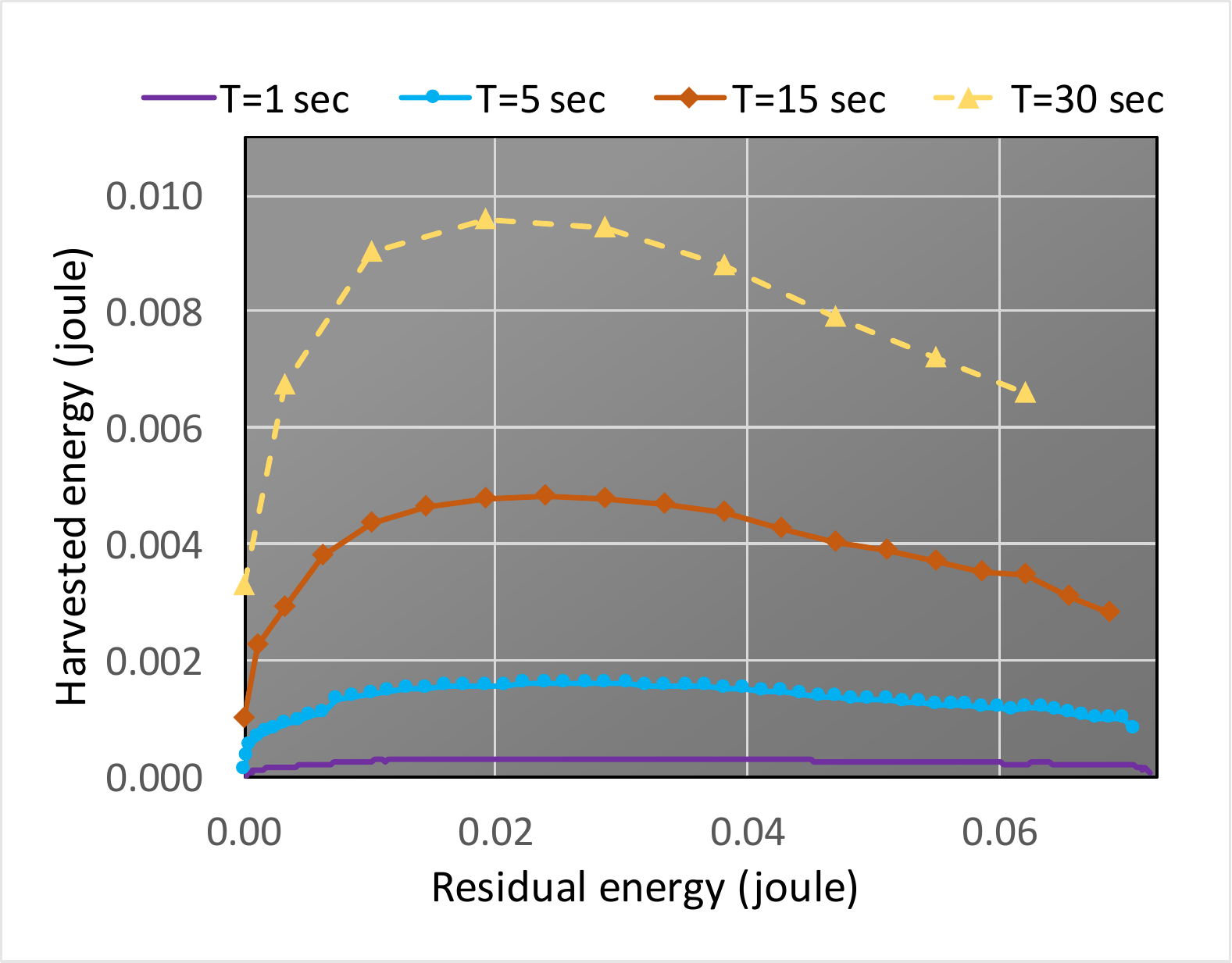}}
  \caption{Nonlinear energy harvesting with respect to the residual energy of a commercial EHN.}\label{fig:Nonlinear}
\end{figure}

Fig.\,\ref{fig:Nonlinear} shows the experimental results that how harvested energy, $E_h$, changes nonlinearly with respect to the residual energy, $E_r$, for different charging time. To be specific, when $E_r$ changes, $E_h$ is not a constant but a concave and non-monotonic function of $E_r$. 
Therefore, it is crucial to retain an appropriate amount of energy to maximize the efficiency of energy harvesting. Since residual energy is determined by data transmission, this means that the power management on EHNs influences harvested energy. To capture this important interplay between the energy harvest and the data transmission, a new feedback-based EH model is proposed, as illustrated in Fig.\,\ref{fig:Mod}(b). In the new model, an energy feedback loop from data transmission to harvested energy is established that captures this important interplay between the process of energy harvest and the power management.


The feedback loop in Fig.\,\ref{fig:Mod}(b) poses a challenge to the design of the optimal strategy for power management. The existing work of power management on EHNs is to seek an energy consumption curve within a fixed energy tunnel so that a pre-defined objective is optimized (e.g., maximizing throughput). In the new energy harvesting model, however, the feasible energy tunnel is not fixed: its bounds are impacted by the residual energy of EHN.This indicates that the energy tunnel, which determines the optimal strategy for power management, is in turn affected by the power management.  In other words, an EHD cannot estimate what amount of energy it can harvest from an energy packet before scheduling its transmissions, and we call it as the \emph{causality of energy harvest}. How to manage the power under the feedback loop effect should be considered carefully.

\section{Challenges on Link Layer}
\label{sec:MAC}

In addition to coordinating the channel access among multiple EHNs, the link layer of an RF-EHSN also needs to schedule the energy harvesting process. When an EHN sends data, neighboring EHNs are able to harvest energy from the overhead signal. As a consequence, the data flow is entangled with the energy flow, which is a unique feature of RF-EHSNs. In this section, we analyze how this feature challenges the implementation of energy and medium access control (EMAC) at the link layer of RF-EHSNs.

\subsection{Efficient Information and Power Transfer}
\label{subsec:Eff}
In a healthcare network, the data generation rates of different biosensors can be different. For example, a sensor monitoring the heartbeat of an animal has much higher sampling rate than the one detecting the presence of pathogens. Due to the heterogeneous traffic loads, EHNs in a network may consume energy at distinct rates. As a result, a node with heavy traffic may not capture sufficient energy from ambient RF signals for extensive data transmission. By contrast, an idle node may harvest superfluous energy that overflows its energy storage gradually, which causes a waste. How to reallocate the energy resource among EHNs is a critical issue.

To balance the energy among EHNs, an available solution is the multi-hop energy transfer. More specifically, given the short distances among nodes, an EHN can harvest substantial amount of energy from its neighbors
~\cite{mishra2015smart}. For instance, the typical stocking density of broilers in Europe is between 11 to 25 birds per square meter. Accordingly, the space among EHNs carried by neighboring chickens is less than 1 feet. If an EHN  transmits data at 0.1\,mW, the neighbors 25\,cm away can harvest 12.7\, nW/cm$^2$ of energy, which is comparable to the energy density of ambient RF signals in a semi-urban area~\cite{pinuela2013ambient}.

Through the multi-hop energy transfer, an EMAC can arrange the EHN with a high residual energy but light traffic to charge its neighbors during the data transmission. However, fast information delivery and efficient energy transfer may not be able to achieve at the same time. It has been studied in \cite{zhang2013mimo} that there exists a trade-off between the maximal data rate and the optimal energy transfer. Furthermore, unlike conventional MAC protocols that consider the signal to unintended receivers as interference degrading the communication reliability, the co-channel noises in RF-EHSNs are beneficial from the viewpoint of energy harvesting. Therefore, the EMAC protocols in RF-EHSNs not only need to avoid collisions among data packets but also seek for effective channel and energy allocations among EHNs for the efficient information delivery and energy transfer.

\subsection{Balance between Energy Cycle and Data Cycle}
\label{subsec:Bal}

Due to the size constraint, the EHN implanted inside animals is commonly equipped with a single antenna switching between energy harvesting (i.e., energy cycle) and data communication (i.e., data cycle). If an EHN is scheduled to receive energy over frequently, it will be less likely to send the data in time, which is unacceptable in applications requiring a real-time health monitoring. In addition, continuous energy harvesting with less consumption leads to a high residual energy and a low charge efficiency, as depicted in Fig.\,\ref{fig:Nonlinear}. However, if an EHN sends packets freely without considering the residual energy, it may fail to harvest enough energy, resulting in a low throughput performance. Therefore, EHNs need to balance the cycles for data transmission and energy reception.

How to schedule the energy cycle and the data cycle for an EHN is not trivial; many factors, such as the length of data queue, residual energy, power density in the environment, and future data arrivals, need to be considered comprehensively. Taking the long data queue as an instance, the EMAC tends to initiate a data cycle when the residual energy is sufficient, thus mitigating the queuing delay. If the residual energy is low and the data queue is short, cumulating energy in the current period for the future data transmission is apparently a wise choice.

In most cases, however, the scenario is not as simple as the above two examples. Take an EHN with a long data queue as an example. If current RF environment has a temporally high energy density, switching to energy cycle may harvest more energy but at the cost of higher latency for data communications. By contrast, continuing on data cycle loses the opportunity to receive a substantial amount of energy but reduces the queuing delay. The balance between energy harvest and data transmission will need a careful calculation. Moreover, the values of residual energy, data queue and energy density in a real application cannot be simply represented by a binary (high or low) but change continuously with time, which makes the coordination between the data cycle and the energy cycle more complicated.

In addition, the network traffic load also affects the scheduling of data communication and energy harvesting. In an RF-EHSN with light traffic, assume the optimal ratio of energy cycles to data cycles is $8\!:\!2$. When the network traffic becomes heavy, the EHN may fail to access the channel in its data cycle to prevent the potential interference with the neighborhood. Consequently, the data cycle arranged for the information transfer is forced into an energy cycle for energy harvesting, which may cause superfluous energy harvest and descended throughput. To tackle this problem, the EMAC  needs to adjust the duty cycle of each EHN adaptively according to the network traffic, thereby maintaining an optimal duty cycle in real-time.

\subsection{Optimal Energy Request}
\label{subsec:Opt}
In some applications like the one illustrated in Fig.\,\ref{fig:scenario}, it is promising to deploy a dedicated energy source (e.g., AP in a closed-access LAHN or femtocell in a 5G network) if the surrounding ambient RF signals are too weak. EHNs carried by animals are allowed to request energy from the associated facilities on demand. The energy replenishment thus can be scheduled based on the energy consumption of each EHN. How an EHN requests energy from the dedicated energy source intelligently to minimize the overall energy consumption while guaranteeing successful data transmission for the efficient healthcare is an interesting problem.

Intuitively, the EHN needs to pay a cost for the transmission of energy requests, thereby producing considerable energy overhead if a node requests over frequently. By contrast, requesting a large amount of energy each time reduces the overhead, but leads to energy inefficiency due to the nonlinear charge characteristic of EHNs, as discussed in Section\,\ref{sec:PHY}. An optimal strategy is necessary for EHNs to request an appropriate amount of energy at the right time. To solve this problem, a path-oriented method is a feasible solution~\cite{luo2017optimal}.

\begin{figure}[htb]
\centerline{\includegraphics[width=8.3cm]{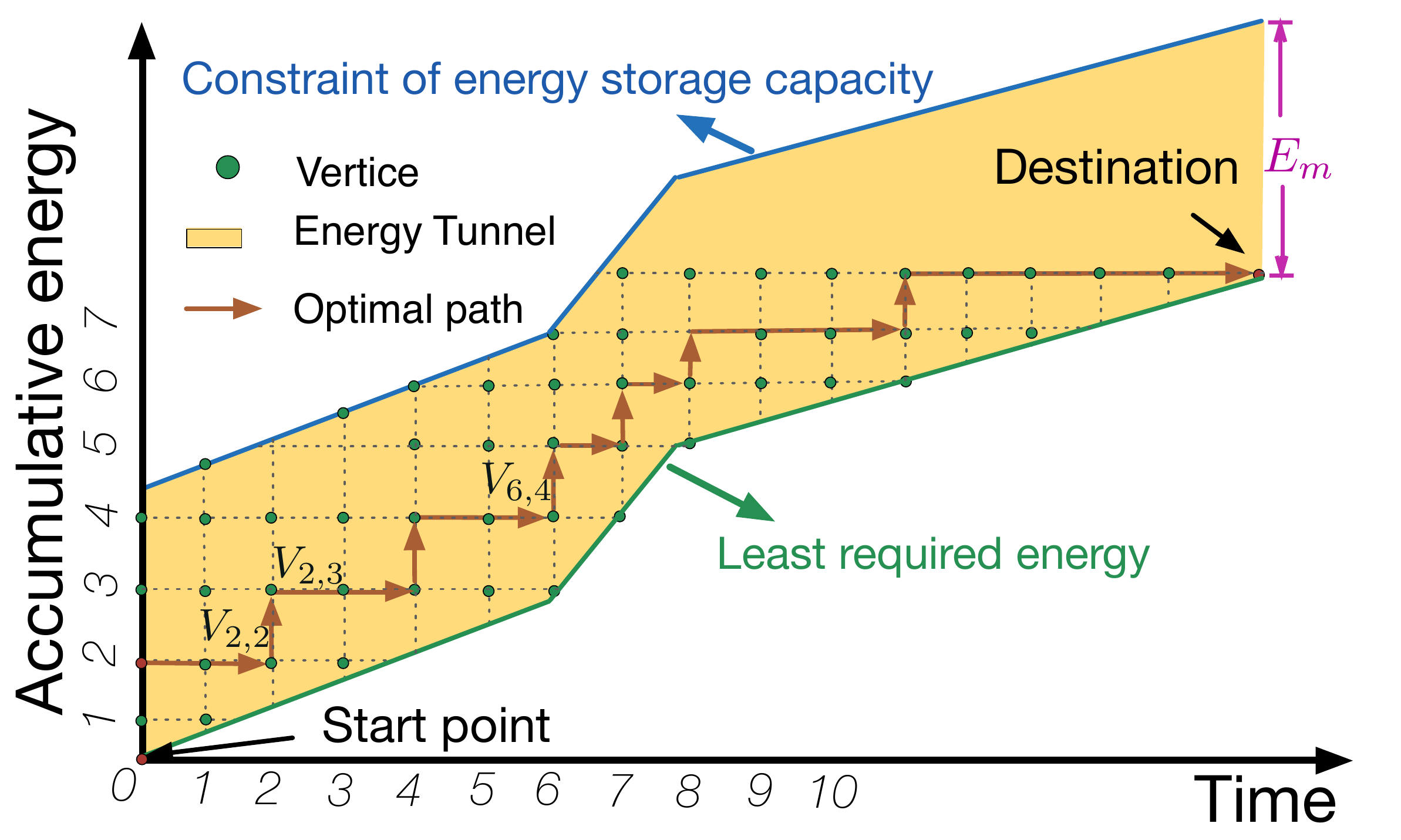}}
  \caption{The shortest path based method for the optimal energy request, where a vertex at coordinates $(i, j)$ is denote by $V_{i, j}$.}\label{fig:ShPath}
\end{figure}

\begin{figure*}[htb]
\centerline{\includegraphics[width=14cm]{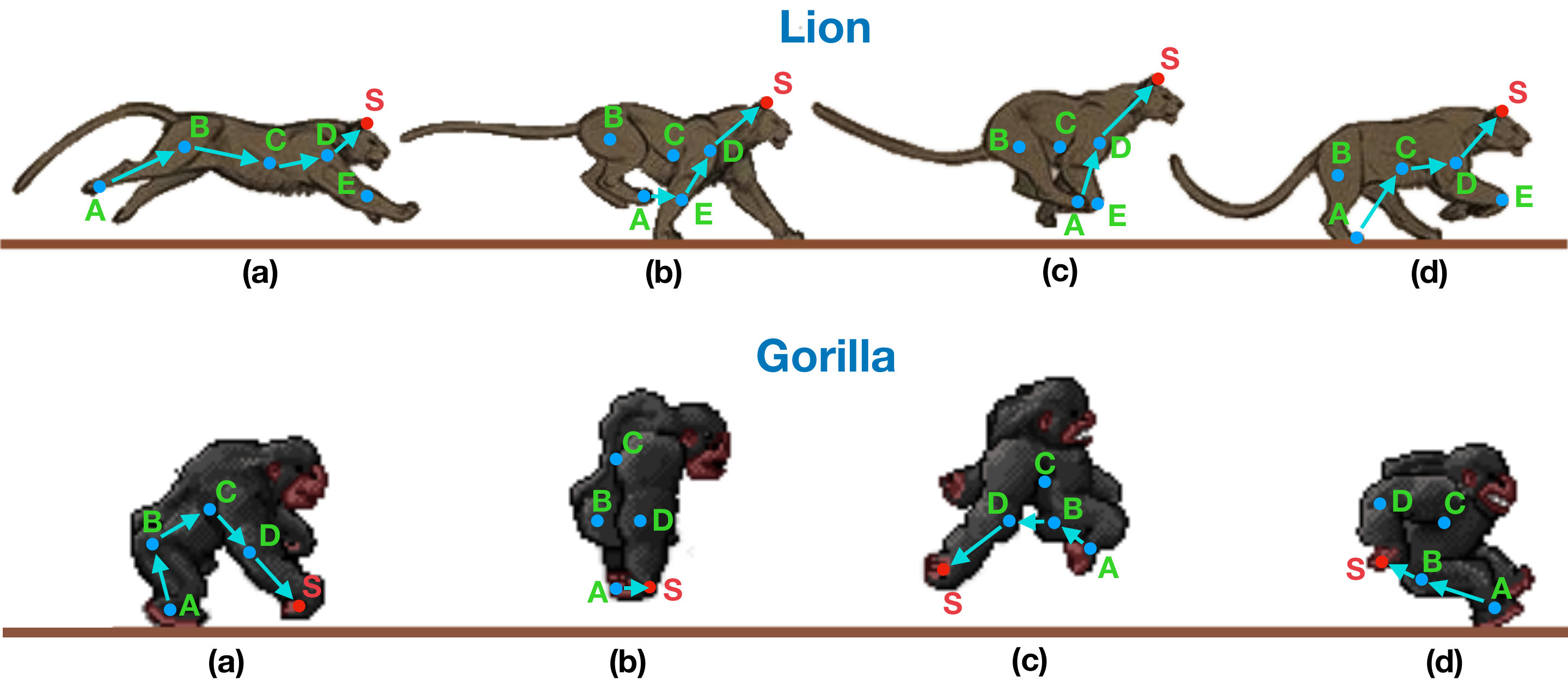}}
  \caption{Topology changes of RF-EHSN resulting from animal movement, where the green dots represent biosensors and the red ones are the sink nodes.}\label{fig:movement}
\end{figure*}

In the path-oriented method, an EHN first calculates the least transmission power it required for a timely delivery of data collected from biosensors. Afterward, an energy tunnel is formed as shown in Fig.\,\ref{fig:ShPath}. The lower bound of the tunnel is determined by the least required energy accumulated with time, and the upper bound is in parallel with but $E_m$ above the lower bound, where $E_m$ is the maximum energy can be stored in an EHN. The accumulation of harvested energy  (red arrow lines in Fig.\,\ref{fig:ShPath}) is subject to the two bounds, otherwise the harvested energy will either be insufficient for a timely data transmission (below the lower bound) or overflow from the energy storage (over the upper bound).

To find the optimal way for energy requesting, the energy tunnel in Fig.\,\ref{fig:ShPath} is divided into multiple grids, forming a graph with a set of ``vertices'' and ``edges''.  The vertical edge indicates an energy replenishment, which generates an associate charging cost at the energy source. The cost consists of two parts: a constant overhead, and a nonlinear charge cost. The latter is determined by both the residual energy of the EHN, which is the vertical distance to the lower bound, and the amount of energy requested, which is the length of the vertical edge. A horizontal edge means no energy charging and generates no cost at the energy source. Eventually,  the optimal energy requesting strategy is converted to find the route between the start point and destination that has the minimum sum-cost along the path. In Fig.\,\ref{fig:ShPath}, the path is not allowed to move backward or downward since the cumulation of  harvested energy increases monotonically with time.

Consequently, the dynamic programming methods like Dijkstra's algorithm can be applied to schedule the optimal energy request. However, such solution can only work in the scenario where the destination is specified initially. The scenarios include the regular health examination, in which the lower bound of energy tunnel is predictable since biosensors collect data periodically from the animal and the EHN transmits the data in a prescheduled pattern. However, in an event-driven based application, such as detecting viruses and pathogens, the presence of an event is random; therefore an EHN cannot arrange its data transmission in advance. In this case, the energy requesting strategy needs to run in an online mode, which is still an open issue.

\section{Challenges at Network Layer}
\label{sec:Routing}

Due to the severe energy constraint on EHNs,  a well-designed multi-hop routing protocol plays a more important role in RF-EHSN than in traditional sensor networks to transmit data reliably and efficiently. In this section, we identify the unique features, challenges and potential solutions of routing design in RF-EHSNs for the healthcare of animals.

\subsection{Dynamic Changes of Network Topology}
\label{subsec:Eff}

RF-EHSNs and mobile ad hoc networks share the feature of highly dynamic topology. However, the topology change in RF-EHSN is caused not only by the mobility of animals carrying biosensors but also by the body movement of an individual and frequent state switching of EHNs.

\textbf{Body Movement:}
Fig.\,\ref{fig:movement} illustrates that in an ad-hoc BAN, how the optimal routing path from biosensor A to sink node S varies in running animals. As shown in the snapshot (a) for a lion, the information at $A$ needs to go through nodes $B$, $C$, and $D$ to reach the destination, $S$. However, after a short period of time, the network topology is changed with the body movement of the lion, as shown in the snapshot (c). At this moment, node $A$ can reach the destination through $D$ directly, which is two hops shorter than that in the snapshot (a). 

An inherent feature of such a dynamic topology revealed in Fig.\,\ref{fig:movement} is the \emph{repetition}. Particularly, different from nodes in a mobile network, which may move randomly, the relative positions of biosensors in an ad-hoc BAN change repeatedly depending on an animal's behavior (e.g., walking, running, and sleeping). For each behavior, the changes of network topology follow a corresponding pattern and are predictable at a certain level. For a routing protocol, the discovery of optimal paths from biosensors to the destination can be fast if the current gesture of an animal is known to the network. The gesture recognition using the three-dimensional accelerometer and inertial sensors has been well studied in recent years \cite{chen2017survey}.

In addition to the repetition, another critical feature of the topology change in Fig.\,\ref{fig:movement} is the species-dependent discrepancy. Intuitively, different animals may have completely different postures for moving and sleeping, which can be observed from the running sequence of a lion and a gorilla in the figure. To minimize the end-to-end delay, the deployment of biosensors and sink node should take the appearance and motion characteristic of animals into account, and the optimal path from an EHN to the sink node can be pre-arranged once the species and the current gesture of an animal are determined through the gesture recognition.

\textbf{Frequent State Switching:} Different from traditional wireless sensor nodes, which consume energy monotonically without battery replenishment, the energy level of an EHSN could even rise after a data transmission and sensing process due to newly harvested energy. Consequently, the sensor node in an EHN will not ``die'' permanently but comes alive after a short break as a perpetual wireless equipment. 
Due to the thin density of the renewable RF energy, a small EHN usually consumes energy much faster than energy harvest rate~\cite{pinuela2013ambient}. In other words, an EHN might fail quickly and lose connections with its neighbors after continuous data transmissions, and then be resurrected after energy harvest. 
The state switch of nodes between live and dead causes frequent and unpredictable changes in the network topology, which poses grand challenges on the routing design. Although there has been a tremendous amount of routing protocols and topology management methods developed for wireless sensor networks~\cite{younis2014topology, pantazis2013energy}, they may not work efficiently in RF-EHSNs due to the \emph{intermittent connections} among nodes. How to sustain the reliable network connectivity for a successive packet delivery is a crucial problem for the routing design in RF-EHSNs. Different from the network carried by a moving animal, the topology change caused by the intermittent connections can even happen in a static network, e.g., ad-hoc BAN carried by a resting or sleeping wildlife, which challenges the reliability of routing protocols.

\subsection{Integration of Energy Flow and Data Flow  }
\label{subsec:Int}

In an RF-EHSN, EHNs can harvest energy from data transmitted by the neighborhood. The amount of harvested energy will vary depending on routes. This adds a new dimension to the innovative routing protocol design by considering the interplay between the energy flow and data flow. For instance, a desired route might be selected from the perspective of total energy harvested from neighboring EHNs along the route.

\begin{figure}[htb]
\centerline{\includegraphics[width=8.8cm]{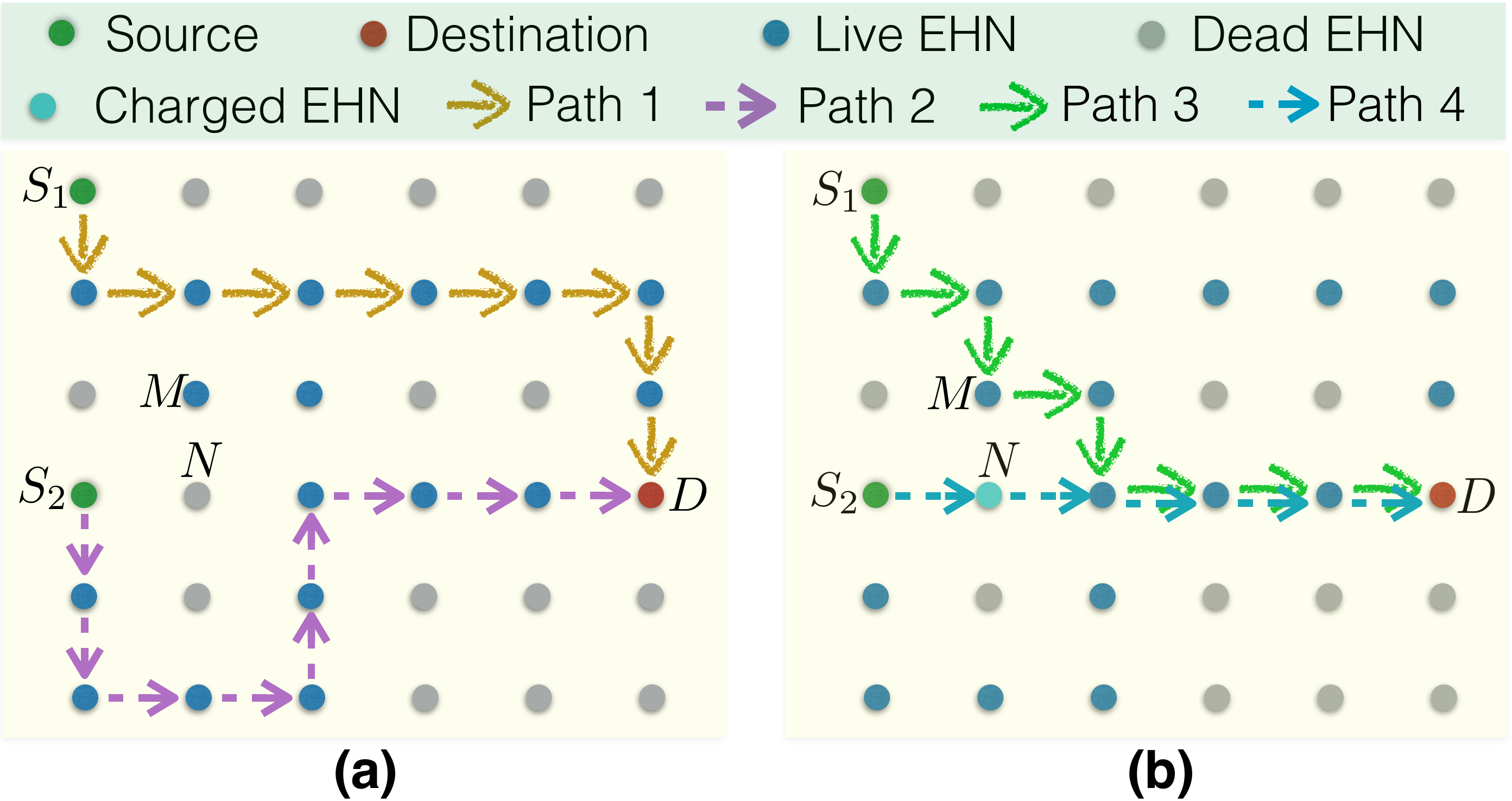}}
  \caption{Integration of energy flow and data flow for the route selection.}\label{fig:EngFlow}
\end{figure}

Fig.\,\ref{fig:EngFlow} illustrates an example on how the energy flow affects the route selection, where two sources, $S_1$ and $S_2$, plan to send data to a common destination, $D$. Assume $S_1$ generates and transmits data prior to $S_2$. From Fig.\,\ref{fig:EngFlow}\,(a), it can be observed that if $S_1$ selects the Path 1 as the route, then $S_2$ will use Path 2 since node $N$ is unavailable. However, if $S_1$ chooses Path 3 as shown in Fig.\,\ref{fig:EngFlow}\,(b), node $N$ will get charged by the data transmission from node $M$, and thus $S_2$ can use Path 4 for information delivery. In Fig.\,\ref{fig:EngFlow}\,(b), the sum of hop count from $S_1$ and $S_2$ to $D$ is $13$, which is four hops less than that in Fig.\,\ref{fig:EngFlow}\,(a). This example provides insights into demands that the energy flow in an RF-EHSN should be integrated with data flow in a routing design.

\section{Conclusions}
\label{sec:Con}

In this article, we give a tutorial on the RF-EHSN for the healthcare of pets, livestock, and wildlife. A three-tier  architecture is proposed to create an online heal monitoring network. Several unique features, such as the interaction between data transmission and energy harvest, the balance issue between a data flow and an energy flow, and dynamic of network topology, are introduced. How those features challenge the design of an RF-EHSN on different layers are discussed to shed light on the resilient RF-EHSN design. We believe RF-EHSN is a promising technology for the health monitoring and disease control of animals.



\begin{thebibliography}{10}
\providecommand{\url}[1]{#1}
\csname url@samestyle\endcsname
\providecommand{\newblock}{\relax}
\providecommand{\bibinfo}[2]{#2}
\providecommand{\BIBentrySTDinterwordspacing}{\spaceskip=0pt\relax}
\providecommand{\BIBentryALTinterwordstretchfactor}{4}
\providecommand{\BIBentryALTinterwordspacing}{\spaceskip=\fontdimen2\font plus
\BIBentryALTinterwordstretchfactor\fontdimen3\font minus
  \fontdimen4\font\relax}
\providecommand{\BIBforeignlanguage}[2]{{%
\expandafter\ifx\csname l@#1\endcsname\relax
\typeout{** WARNING: IEEEtran.bst: No hyphenation pattern has been}%
\typeout{** loaded for the language `#1'. Using the pattern for}%
\typeout{** the default language instead.}%
\else
\language=\csname l@#1\endcsname
\fi
#2}}
\providecommand{\BIBdecl}{\relax}
\BIBdecl

\bibitem{shaikh2016energy}
F.~K. Shaikh and S.~Zeadally, ``{Energy harvesting in wireless sensor networks:
  a comprehensive review},'' \emph{Renewable and Sustainable Energy Reviews},
  vol.~55, pp. 1041--1054, 2016.

\bibitem{hu2017wireless}
F.~Hu, X.~Liu, M.~Shao, D.~Sui, and L.~Wang, ``{Wireless Energy and Information
  Transfer in WBAN: an overview},'' \emph{IEEE Network}, vol.~31, no.~3, pp.
  90--96, 2017.

\bibitem{neethirajan2017recent}
S.~Neethirajan, ``Recent advances in wearable sensors for animal health
  management,'' \emph{Sensing and Bio-Sensing Research}, vol.~12, pp. 15--29,
  2017.

\bibitem{salayma2017wireless}
M.~Salayma, A.~Al-Dubai, I.~Romdhani, and Y.~Nasser, ``{Wireless body area
  network (WBAN): a survey on reliability, fault tolerance, and technologies
  coexistence},'' \emph{ACM Computing Surveys}, vol.~50, no.~1, p.~3, 2017.

\bibitem{sun2016edgeiot}
X.~Sun and N.~Ansari, ``{EdgeIoT: mobile edge computing for the internet of
  things},'' \emph{IEEE Communications Magazine}, vol.~54, no.~12, pp. 22--29,
  2016.

\bibitem{kurup2012body}
D.~Kurup, W.~Joseph, G.~Vermeeren, and L.~Martens, ``In-body path loss model
  for homogeneous human tissues,'' \emph{IEEE Transactions on Electromagnetic
  Compatibility}, vol.~54, no.~3, pp. 556--564, 2012.

\bibitem{pinuela2013ambient}
M.~Pinuela, P.~D. Mitcheson, and S.~Lucyszyn, ``{Ambient RF energy harvesting
  in urban and semi-urban environments},'' \emph{IEEE Transactions on Microwave
  Theory and Techniques}, vol.~61, no.~7, pp. 2715--2726, 2013.

\bibitem{buzzi2016survey}
S.~Buzzi, I.~Chih-Lin, T.~E. Klein, H.~V. Poor, C.~Yang, and A.~Zappone, ``{A
  survey of energy-efficient techniques for 5G networks and challenges
  ahead},'' \emph{IEEE Journal on Selected Areas in Communications}, vol.~34,
  no.~4, pp. 697--709, 2016.

\bibitem{ulukus2015energy}
S.~Ulukus, A.~Yener, E.~Erkip, O.~Simeone, M.~Zorzi, P.~Grover, and K.~Huang,
  ``Energy harvesting wireless communications: a review of recent advances,''
  \emph{IEEE Journal on Selected Areas in Communications}, vol.~33, no.~3, pp.
  360--381, 2015.

\bibitem{boshkovska2015practical}
E.~Boshkovska, D.~W.~K. Ng, N.~Zlatanov, and R.~Schober, ``{Practical nonlinear
  energy harvesting model and resource allocation for SWIPT systems},''
  \emph{IEEE Communications Letters}, vol.~19, no.~12, pp. 2082--2085, 2015.

\bibitem{biason2016effects}
A.~Biason and M.~Zorzi, ``On the effects of battery imperfections in an energy
  harvesting device,'' in \emph{Computing, Networking and Communications
  (ICNC), 2016 International Conference on}.\hskip 1em plus 0.5em minus
  0.4em\relax IEEE, 2016, pp. 1--7.

\bibitem{mishra2015smart}
D.~Mishra, S.~De, S.~Jana, S.~Basagni, K.~Chowdhury, and W.~Heinzelman,
  ``{Smart RF energy harvesting communications: challenges and
  opportunities},'' \emph{IEEE Communications Magazine}, vol.~53, no.~4, pp.
  70--78, 2015.

\bibitem{zhang2013mimo}
R.~Zhang and C.~K. Ho, ``{MIMO broadcasting for simultaneous wireless
  information and power transfer},'' \emph{IEEE Transactions on Wireless
  Communications}, vol.~12, no.~5, pp. 1989--2001, 2013.

\bibitem{luo2017optimal}
Y.~Luo, L.~Pu, Y.~Zhao, and G.~Wang, ``{Optimal energy requesting strategy for
  RF energy harvesting wireless communications},'' in \emph{Proceedings of
  INFOCOM}.\hskip 1em plus 0.5em minus 0.4em\relax IEEE, 2017, pp. 1--9.

\bibitem{chen2017survey}
C.~Chen, R.~Jafari, and N.~Kehtarnavaz, ``A survey of depth and inertial sensor
  fusion for human action recognition,'' \emph{Multimedia Tools and
  Applications}, vol.~76, no.~3, pp. 4405--4425, 2017.

\bibitem{younis2014topology}
M.~Younis, I.~F. Senturk, K.~Akkaya, S.~Lee, and F.~Senel, ``{Topology
  management techniques for tolerating node failures in wireless sensor
  networks: a survey},'' \emph{Computer Networks}, vol.~58, pp. 254--283, 2014.

\bibitem{pantazis2013energy}
N.~A. Pantazis, S.~A. Nikolidakis, and D.~D. Vergados, ``Energy-efficient
  routing protocols in wireless sensor networks: a survey,'' \emph{IEEE
  Communications surveys \& tutorials}, vol.~15, no.~2, pp. 551--591, 2013.

\end{thebibliography}

\end{document}